\documentclass[aps,prd,preprintnumbers,superscriptaddress,nofootinbib,twocolumn]{revtex4-2}
\usepackage{graphicx}
\usepackage{bm,latexsym,amsmath,amssymb,amsfonts,mathrsfs}
\usepackage{color}
\input{colordvi.tex}
\usepackage{enumitem} 
\usepackage{dirtytalk}
\usepackage{mathtools,mathbbol}
\usepackage{graphicx,hyperref,xcolor}
\usepackage{lipsum}
\usepackage{float}

\def\ba{\begin{array}}
\def\ea{\end{array}} 
\def\bea{\begin{eqnarray}}
\def\eea{\end{eqnarray}}
\def\beq{\begin{equation}}
\def\eeq{\end{equation}}
\def\ben{\begin{enumerate}}
\def\een{\end{enumerate}}

\def\brr{\begin{array}}
\def\err{\end{array}}

\usepackage[normalem]{ulem}

\begin{document}


\title{Cosmological Bounce Relics:\\ Black Holes, Gravitational Waves, and Dark Matter}
\author{Enrique Gazta\~naga}

\affiliation{Institute of Cosmology \& Gravitation, University of Portsmouth, Portsmouth PO1 3FX, United Kingdom}
\affiliation{Institute of Space Sciences (ICE, CSIC), 08193 Barcelona, Spain}
\affiliation{Institut d'Estudis Espacials de Catalunya (IEEC), 08034 Barcelona, Spain}

\begin{abstract}
We propose a new mechanism for generating cosmological relics—black holes, gravitational waves (GWs), and possibly dark matter (DM)—in a bouncing Universe. Relics arise through two channels: (i) compact objects and GWs produced during pre-bounce collapse that remain super-horizon and re-enter after the bounce, and (ii) dark-matter halos formed during collapse that exit the horizon and collapse into black holes upon re-entry. Unlike inflationary primordial black holes, these relic black holes originate from nonlinear structure formation during collapse. We derive the particle-horizon and horizon-crossing conditions in bouncing cosmology and show that perturbations or compact objects larger than $\sim 90$ m survive the bounce. The resulting population of relic black holes and GWs spans a wide mass range, offering a unified origin for dark matter, gravitational-wave backgrounds, and the early growth of supermassive black holes and galaxies.

\end{abstract}

\maketitle

 \section{Introduction}
The nature of dark matter remains one of the most profound mysteries in cosmology, with no direct detection of a candidate particle to date. Numerous possibilities arise from extensions of the Standard Model of particle physics. These include \textit{WIMPs} (weakly interacting massive particles) \cite{Bertone2005}, \textit{axions}—extremely light bosons proposed to solve the strong-CP problem in QCD \cite{Preskill1983}, \textit{sterile neutrinos}—hypothetical particles that interact only via gravity and weak mixing with active species \cite{Dodelson1994}, and \textit{fuzzy dark matter}—ultralight scalars whose kiloparsec-scale de~Broglie wavelength smooths galactic cores and potentially resolves small-scale structure issues \cite{Hu2000}. A different class of candidates is comprised of \textit{primordial black holes} (PBHs), which are formed in the early Universe from the direct collapse of rare overdensities \cite{Carr1974,Carr1975}. Each of these candidates carries distinct astrophysical signatures, and ongoing experiments seek their detection either directly (via scattering in underground detectors), indirectly (via decay or annihilation products), or gravitationally (via lensing, clustering, or dynamical effects).

Primordial black holes face two major challenges as dark matter candidates. First, producing them during inflation is difficult: for scale-invariant Gaussian fluctuations, only extremely rare overdensities exceed the threshold for collapse at horizon entry. Even if PBHs do form in sufficient numbers, their mass spectrum must evade numerous observational constraints \cite{GarciaBellido2017a,GarciaBellido2017b}.

Another remarkable discovery in astrophysics is the tight empirical correlation between galaxies and their central supermassive black holes (SMBHs). Almost every massive galaxy hosts an SMBH, whose mass correlates strongly with the stellar bulge velocity dispersion (the $M$--$\sigma$ relation) and with the host's total stellar mass  \cite{Ferrarese2000,Gebhardt2000}. These relationships suggest that black holes and galaxies co-evolve, regulating each other's growth. Two broad interpretations frame this connection:
\begin{itemize}[leftmargin=*]
    \item \textit{Nurture:} The correlations result from co-evolution. As galaxies grow by accreting gas and merging, their SMBHs also accrete and inject energy into their surroundings. This active galactic nucleus (AGN) feedback can regulate star formation, drive gas outflows, and establish a self-consistent scaling between stellar and black-hole mass.
    \item \textit{Nature:} The correlations instead reflect common initial conditions. In this view, some black holes may have formed even before galaxies—perhaps via relativistic collapse or a cosmological bounce—and served as seeds around which galaxies assembled. Galaxy properties would then be inherited from pre-existing black holes rather than determined by their subsequent evolution.
\end{itemize}
Current evidence suggests that both processes are at play. Simulations and observations support the feedback scenario, but it remains unclear whether feedback alone can explain the tightness of the relations, particularly at high redshifts. The discovery of massive quasars less than a billion years after the Big Bang intensifies this tension: growing such large SMBHs so quickly via standard accretion and mergers remains a serious challenge. Discovering primordial or otherwise relic black holes could shift this paradigm, pointing toward a gravitational (rather than baryonic) origin for the seeds of cosmic structure.

Could all dark matter be made of baryons? Constraints from Big Bang nucleosynthesis and the cosmic microwave background (CMB) apply only to baryons that remained diffuse—e.g., protons, electrons, and nuclei in thermal equilibrium with radiation in the early Universe. Baryonic material could still constitute dark matter if confined within compact objects. For instance, cosmological $N$-body simulations (e.g. \cite{Davis1985,Fosalba2008,Fosalba2015}) treat each simulation particle as a massive object—sometimes comparable in mass to an SMBH—that interacts solely via gravity. Astronomical compact objects such as black holes or neutron stars behave similarly: effectively collisionless and gravitationally interacting.

In this paper, we propose a new class of dark matter candidate: \textit{bounce dark matter} (BDM). It consists of compact objects—such as black holes or neutron stars—produced during a cosmological bounce. For concreteness, we focus on the \textit{black hole universe} (BHU) model (see Appendix~\ref{sec:BHU}), which operates entirely within classical general relativity. However, the same mechanism can arise in other bouncing cosmologies involving modified gravity theories~\cite{Novello2008,Poplawski2016,Brandenberger2017}. The key physical ingredient is the behavior of perturbations near the bounce and their relation to horizon crossing (see Section~\ref{sec:horizoncrossing}).

This paper is organized as follows. Appendices~\ref{sec:perturbation} and \ref{sec:BHU} introduce relativistic perturbations in a collapsing FLRW* cloud and provide a self-contained derivation of the bounce solution, highlighting the role of curvature and energy conditions. Section~\ref{sec:virialization} reviews halo virialization, which sets the stage for compact object formation. Section~\ref{sec:Bounce_Solution} describes the background evolution in a bouncing cosmology, naturally incorporating both inflation and late-time acceleration. Section~\ref{sec:horizoncrossing} examines horizon crossing, a key mechanism that determines whether fluctuations or black holes survive the bounce. Finally, in Section~\ref{sec:BDM}, we present the Bounce Dark Matter scenario and its relation to PBHs and gravitational waves, outlining two formation channels and the expected population of relics from a cosmological bounce.

\begin{figure}
\includegraphics[width=0.9\columnwidth]{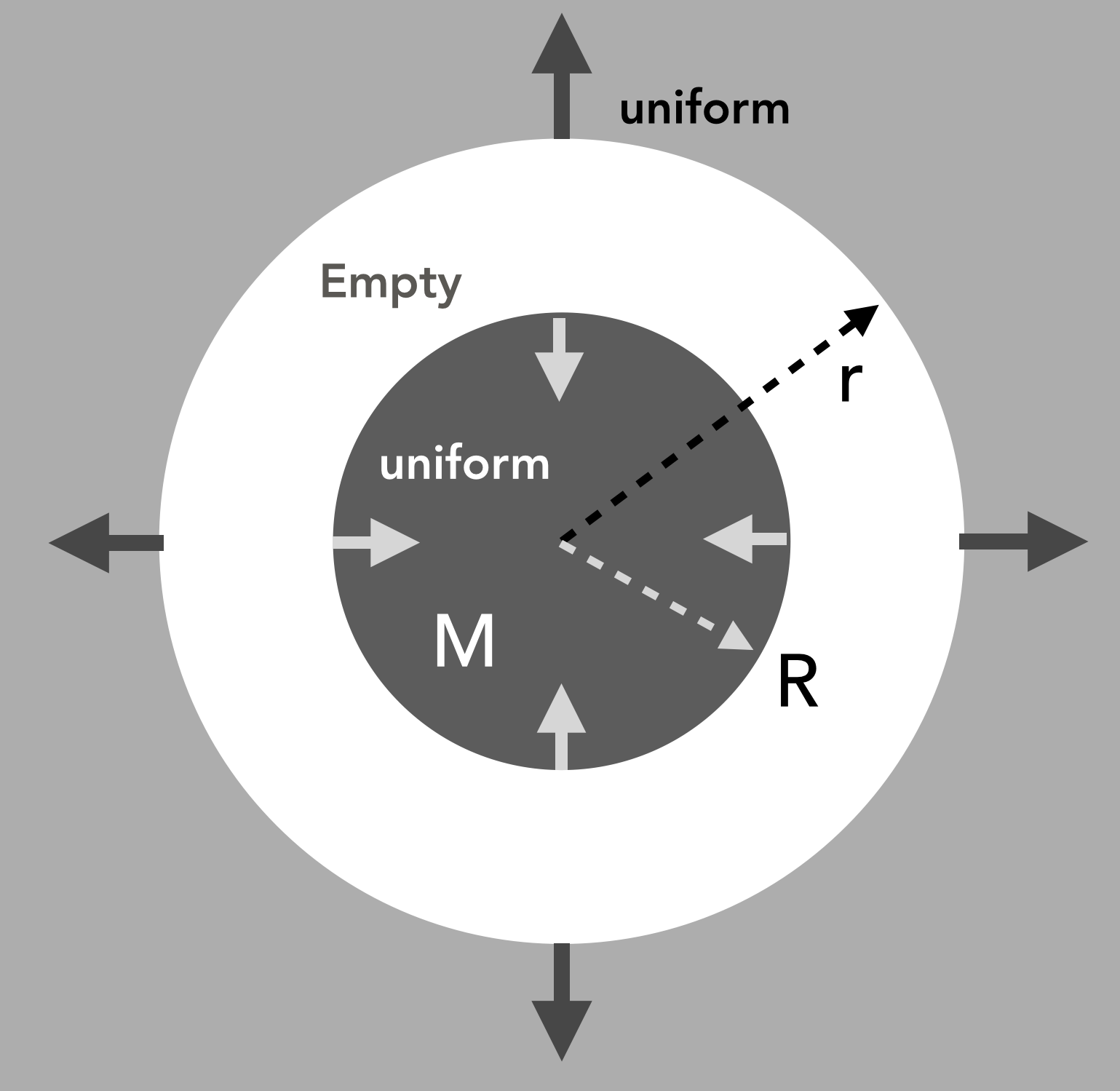}
\caption{\textbf{Configuration of a spherical collapse perturbation.} We consider three spherically symmetric regions: (1) an outer background with radius $>r$ and mean density $\bar\rho$; (2) an inner region of radius $R<r$ with higher density $\rho = \bar\rho(1+\delta)$; and (3) an intermediate near-vacuum layer separating them. This can represent a collapsing overdensity of size $R$ within a larger universe of size $r$, or analogously, our finite Universe of radius $R$ embedded in a larger space.}
\label{fig:collapse}
\end{figure}

\section{Dark Matter Halo Virialization}
\label{sec:virialization}

Gravitational collapse of collisionless matter—such as dark matter—does not continue indefinitely. As matter collapses, particle orbits cross and exchange energy through a process known as \textit{violent relaxation}. This redistributes kinetic energy and builds up an effective pressure, ultimately halting the collapse. The result is a quasi-equilibrium configuration: a virialized dark matter halo.

The outcome depends strongly on the matter content. Baryonic gas can radiate energy and cool, allowing continued collapse into compact structures such as stars or disks \cite{White1978}. Cold dark matter (CDM), by contrast, lacks efficient cooling and therefore virializes at relatively low densities. Its collapse is rapid but stabilizes into halos governed primarily by gravity.

A virialized halo is characterized by a \textit{virial radius} $R_V$ (its physical size) and a \textit{virial velocity} $\sigma_V$ (the one-dimensional velocity dispersion of its particles). For a system in virial equilibrium (see \cite{2017JCAP...10..040P} and references therein),
\begin{equation}
2K_V + \Phi_V = 0 \quad \Rightarrow \quad \sigma_V^2 = 2K_V = \frac{GM}{R_V} \propto M^{2/3},
\end{equation}
where $K_V$ and $\Phi_V$ are the kinetic and gravitational potential energies.

A spherical top-hat perturbation collapsing in an Einstein–de Sitter (EdS) universe provides a useful benchmark: it virializes at an overdensity of order $18\pi^2 \approx 178$ relative to the cosmic mean at that time. Consider a spherical region of radius $R$ with density $\rho = \bar\rho (1 + \delta)$, initially only slightly above the cosmic mean $\bar\rho$ (see Fig.~\ref{fig:collapse}). To conserve mass,
\begin{equation}
M = \frac{4\pi}{3} r^3 \bar\rho = \frac{4\pi}{3} R^3 \bar\rho (1 + \delta),
\end{equation}
so that the \emph{virial overdensity} becomes
\begin{equation}
\Delta_V = \frac{\rho_V}{\bar\rho} = 1 + \delta_V = \frac{2GM}{H^2 R_V^3} = 18\pi^2 \simeq 178.
\end{equation}

This result implies that CDM halos collapse into densities about 200 times the background value. Numerical simulations and analytic models confirm that the virial overdensity is nearly independent of halo mass, epoch, or initial conditions. From this, one can estimate the halo mass from its virial radius:
\begin{equation}
M \simeq 10^{11} h^{-1} M_\odot \left(\frac{R_V}{120\,h^{-1}\mathrm{kpc}}\right)^3 \left(\frac{\Omega_m}{0.3}\right)(1+z_V)^3.
\label{eq:Mhalo}
\end{equation}

In a BHU bounce scenario, halos that reach this overdensity before the bounce will collapse into black holes upon horizon re-entry.

Once a CDM halo forms, baryons accrete into its potential well, cool, and condense into stars and galaxies \cite{White1978}. Because primordial fluctuations are approximately scale-invariant, structure formation proceeds hierarchically: small halos form first, later merging into larger systems and giving rise to galaxies, clusters, and the cosmic web.

\subsection{Press–Schechter formalism}
The abundance of collapsed objects as a function of mass (the halo mass function) captures the hierarchical assembly of structure. In the simplest analytic model \cite{Press1974}, small Gaussian fluctuations grow via gravity until they exceed a critical overdensity $\delta_c \approx 1.686$ and decouple from the Hubble flow to collapse. Assuming spherical symmetry and Gaussian initial conditions (as predicted by inflation), one can estimate the comoving number density of halos of mass $M$ by counting regions above this threshold. Importantly, because collapse occurs in the matter-dominated era at relatively low densities, it is halted by orbit crossing and violent relaxation, forming an extended halo rather than a singular point.

The Press–Schechter approach \cite{Press1974} and its extensions \cite{Bardeen1986,Bond1991,Sheth1999} predict that high-mass halos are exponentially rare. For a nearly scale-invariant initial spectrum, the halo mass function at the high-mass end approximately follows
\begin{equation}
\frac{dn}{dM} \;\approx\; \frac{1}{\sqrt{\pi}}\left(\frac{M}{M_*}\right)^{1/2} 
\exp\!\Big(-\frac{M}{M_*}\Big)\; \frac{\bar\rho}{M^2},
\label{eq:PS-massfn}
\end{equation}
where $M_*$ is the characteristic mass scale where fluctuations become nonlinear. For general spectral index $n$, this shape is modified only by a slowly varying prefactor \cite{Sheth1999}. Thus, extremely massive halos are rare but allowed.

\section{A Cosmological Bounce Solution}
\label{sec:Bounce_Solution}

As the density in a closed FLRW collapse approaches a critical value, new physics can intervene to prevent a singularity. In the BHU scenario (see Appendix \ref{sec:BHU}), the equation of state becomes extremely stiff at high densities, eventually approaching $p \approx -\rho$ as the energy density asymptotically saturates at a ground-state value $\rho_G$ of constant energy-density (where $\dot \rho = 0$ in Eq.~\ref{eq:ddota}). At this point, the null energy condition is saturated ($T_{\mu\nu}k^\mu k^\nu \to 0$), allowing the collapse to halt and reverse into expansion. In other words, a \textit{gravitational bounce} occurs: the scale factor reaches a minimum $a_B$ and then begins to grow. Because the spatial slices are closed ($k=+1$), one of the key assumptions of the Penrose singularity theorem—the existence of a non-compact Cauchy surface—is violated, helping the system to evade a singularity despite undergoing gravitational collapse \cite{Penrose1965}.

The post (and pre)-bounce expansion naturally includes an inflationary phase. In the BHU solution, the scale factor near the bounce behaves approximately as $a(\tau) \propto \cosh(\tau/R_B)$, where $R_B$ is a characteristic timescale of order the “bounce radius.” For $\tau > 0$, this yields exponential expansion ($a \propto e^{\tau/R_B}$), which drives a period of cosmic inflation, solving the horizon problem and diluting unwanted relics. The inflationary phase ends when the fluid transitions from vacuum domination ($p \approx -\rho$) to matter/radiation domination. Similarly, the late-time accelerated expansion of our Universe (often attributed to a cosmological constant $\Lambda$ or dark energy) can be reinterpreted in this picture as a consequence of the finite mass and size of the FLRW patch: effectively, $\Lambda$ is related to the Schwarzschild radius of the finite collapsing cloud \cite{gaztanaga:bhu1}. We refer the reader to Appendix \ref{sec:BHU} for further details of this correspondence. 

For our purposes, the crucial consequence of the bounce is that it provides a time-symmetric cosmology with both a collapsing pre-bounce phase and an expanding post-bounce phase. Structure can grow during collapse and later influence the expanding universe. In particular, perturbations that exit the horizon before or during the bounce can later re-enter, potentially seeding gravitational structures after the bounce.

\subsection{Bounce Inflation and Particle Horizon}

We consider a phenomenological bounce model in which the energy density evolves smoothly from a matter-dominated scaling, $\rho \propto a^{-3}$, to an asymptotic quantum ground state of nearly constant density, $\rho \simeq \rho_G$ (see \cite{Gaztanaga2025} or Appendix \ref{sec:BHU}). 
A convenient interpolation is
\begin{equation}
\rho = \frac{\rho_G}{1 + [(a - a_B)/a_G]^3},
\end{equation}
where $a_B$ denotes the bounce scale factor, and $a_G \gg a_B$ characterises the transition scale at which the system exits (during expansion) or enters (during collapse) the bounce-inflationary regime, approaching standard matter domination.

With this ansatz, the Hubble rate can be written as
\begin{equation}
H^2 = \frac{a^2 - a_B^2 f(a)}{a^2 R_B^2 f(a)}; \quad 
f(a) = 1 + \left(\frac{a - a_B}{a_G}\right)^3,
\label{eq:Hubble-mod-anstz}
\end{equation}
where $f(a)\ge 2$ signals the onset of the matter-dominated phase for $a \gtrsim a_G$.
Solving this numerically yields the effective equation of state:
$P = -\rho^2/\rho_G$, corresponding to a generalized Chaplygin gas with $\alpha = 2$ (see \cite{Gaztanaga2025,2004PhRvD..69b3004M} and references therein).

\begin{figure*}
\includegraphics[width=1.5\columnwidth]{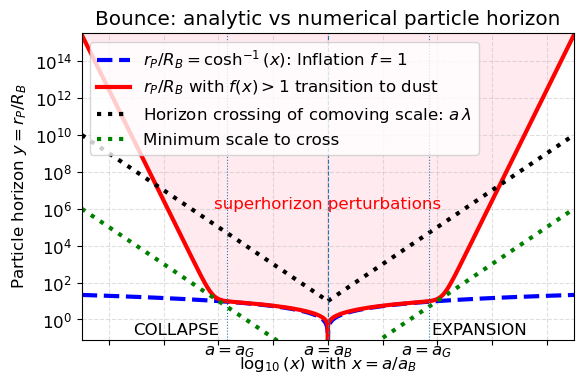}
\caption{\textbf{Particle horizon versus perturbation scale in a bounce cosmology.} The blue dashed curve shows the comoving particle horizon in a bounce scenario with an analytic $\cosh$-inflationary phase; the red curve includes transition from matter domination to the stiff ground state. The black dotted line shows the physical scale $a\,\lambda$ of a perturbation. The green dotted line marks the minimum scale that becomes superhorizon (Eqs.~\ref{eq:lambda_min}–\ref{eq:rHmin}). The pink region shows times when the mode is superhorizon, while the white region denotes subhorizon scales. During collapse (left to center), the horizon shrinks and all $\lambda > 90$ m modes become superhorizon; after the bounce (center to right), the horizon grows and such modes re-enter.}
\label{fig:Horiozon-crossing}
\end{figure*}

The particle horizon (in units of $c=1$) is defined as
\begin{equation}
r_P = a \int \frac{d\tau}{a} = \int_{a_B}^{a} \frac{da}{H a^2}.
\label{eq:rP}
\end{equation}
During the bounce phase ($f \simeq 1$, this yields the analytic solution:
\begin{equation}
r_P = R_B \cosh\left(\frac{a}{a_B}\right) \sinh^{-1}\left(\frac{a}{a_B}\right)  \propto \tau.
\end{equation}
where $\tau= R_B \cosh^{-1}(a/a_B)$ is the time left to the bounce or the time after the bounce: $a = a_B \cosh{\tau/R_B}$.
This is shown as a dashed blue line in Fig.~\ref{fig:Horiozon-crossing} as a function of $\log_{10}(x)$, where $x = a/a_B$. The red line corresponds to the particle horizon $r_P$ (in units of $R_B$) numerically integrated in Eq.~\ref{eq:rP} using $H$ from Eq.~\ref{eq:Hubble-mod-anstz}. For $a \gg a_G$ the universe becomes matter dominated, while $a_G > a > a_B$ corresponds to bounce inflation.

This determines the causal horizon across the bounce. Perturbations with physical size $a\lambda > r_P$ become superhorizon (when crossing inside the pink region) and remain frozen (black dotted line in Fig.~\ref{fig:Horiozon-crossing}). Only those wavelengths larger than the particle horizon at $a \simeq a_G$ survive the bounce:
\begin{equation}
a \lambda_{\min} \simeq r_P(a_G),
\label{eq:lambda_min}
\end{equation}
This scale sets the \emph{minimum relic mass} $M_{\min}$ that can survive as a bounce black hole, where $M_{\min}$ can be estimated from Eq.~\ref{eq:Mhalo} using $R_V = \lambda_{\min}$.

During cold collapse (neglecting pressure and curvature), the GR field equations give
\begin{equation}
\rho = \frac{\tau^{-2}}{6\pi G}
\simeq 3.97 \times 10^{-13}\, \frac{\rm M_{\odot}}{\text{km}^3} \left[\frac{\tau}{\text{s}}\right]^{-2}.
\label{eq:rho}
\end{equation}
At such low densities, thermal pressure $P$ and temperature $T$ are negligible, as collapse timescales are faster than those of interactions between neutral particles. The cold collapse proceeds inside the BH event horizon. Even 1 second before the potential BH singularity, the density is below nuclear saturation density (SD) typical of atomic nuclei or neutron stars (NS):
\begin{equation}
\rho(\tau=1\,\text{s}) \ll \rho_{\rm NS} \simeq \rho_{\rm SD} \simeq 1.4 \times 10^{-4}\, \frac{\rm M_{\odot}}{\text{km}^{3}}.
\label{eq:NS}
\end{equation}
For a total mass $M = 5\times 10^{22}\, \rm M_\odot$, and assuming the ground state $\rho_G > \rho_{\rm SD}$, we find
$R_{\rm G} < r_{\rm SD} = 4.4 \times 10^{8}\, \text{km} \simeq 1.43 \times 10^{-14}\, \rm Gpc$. From Eq.\ref{eq:a_B},
the corresponding Hubble horizon (in units of $c=1$) is:
\begin{equation}
\frac{1}{H} = R_B =  \sqrt{\frac{3}{8\pi G \rho_G}} < 90\,\text{m}.
\label{eq:rHmin}
\end{equation}
Thus, all virialized halos, compact objects, perturbations or gravitational waves with comoving scale $\lambda > 90$ m can become superhorizon and survive the bounce as relics. DM halos and compact objects will collapse into relic black holes upon horizon re-entry, while pre-existing GWs, perturbations and BHs transition smoothly from collapse to expansion.

\section{Horizon Crossing}
\label{sec:horizoncrossing}

Perturbations evolve differently depending on whether their wavelength is larger or smaller than the particle horizon. This distinction between \emph{superhorizon} and \emph{subhorizon} modes is central to relativistic perturbation theory and structure formation \cite{Mukhanov:1988jd,Bernardeau}.

When a perturbation lies on scales larger than the horizon, causal forces cannot act across it. Pressure gradients, sound waves, and gravity have no time to rearrange the overdensity, so the curvature perturbation remains essentially constant (“frozen”). The density contrast similarly retains its primordial value, apart from a small decaying gradient term. In effect, superhorizon fluctuations preserve their initial structure until horizon re-entry.

Once a mode enters the horizon, causal processes can act. In the radiation era ($c_s^2 = 1/3$), perturbations oscillate as acoustic waves and the gravitational potential $\Phi$ decays. In the matter era ($c_s^2 \approx 0$), pressure becomes negligible, $\Phi$ remains nearly constant, and density perturbations grow roughly as $\delta \propto a(\tau)$. This transition underlies many cosmological features: suppression of small-scale power during radiation domination, acoustic peaks in the CMB and LSS, and the onset of galaxy formation after matter–radiation equality.

\subsection{What does it mean to “cross the horizon”?}
In an expanding universe, the particle horizon $r_P$ defines the maximum comoving distance over which causal influences can propagate. Regions separated by more than $r_P$ evolve independently, as if causally disconnected.

A mode with $\lambda \gg r_P$ is termed \textit{superhorizon}. No causal process can act across it, so its shape and amplitude are preserved. This explains why cosmic structure seeded by early-universe processes remains coherent. During inflation, comoving horizon length grows slowly while physical scales expand rapidly, causing modes to exit the horizon. After inflation, the horizon begins to grow and these modes eventually re-enter. Structure formation can only proceed after horizon re-entry, when microphysics (e.g., gravity, pressure) can act on the perturbation.

This cycle of horizon exit and re-entry is crucial for establishing initial conditions in standard cosmology—and it plays an equally central role in bounce scenarios. It allows structures formed during pre-bounce collapse to survive and influence the post-bounce universe.

\section{Bounce DM and GW relics}
\label{sec:BDM}

Astrophysical black holes need not form solely from the deaths of massive stars. Long before galaxies emerged---even before recombination---the early universe could have produced a population of black holes through alternative mechanisms. Such black holes, if present, would carry twofold significance: they would be natural dark matter candidates, and they could provide the seeds for building SMBHs, aiding early galaxy formation.

We identify two distinct pathways by which black holes might arise prior to galaxy formation:

\begin{enumerate}[leftmargin=*, labelindent=0pt, labelsep=0.5em]
    \item \textit{Standard Primordial black holes (PBHs):} These form during the radiation-dominated era from rare, large primordial density fluctuations \cite{Zeldovich1967,Carr1974}. Mathematically, PBH abundance calculations often resemble the Press–Schechter approach, using a Gaussian field, a smoothing scale, and a collapse threshold. This resemblance, however, is superficial. The physics of PBH formation is entirely different from the slow, nonrelativistic collapse that produces ordinary dark matter halos.

    In the Press–Schechter picture, structures form during the matter-dominated era. Densities are low, and collapse proceeds gradually. Orbit crossing leads to violent relaxation, producing virialized halos. The collapse halts far outside the Schwarzschild radius, so no black hole forms.

    PBHs, in contrast, originate in a radiation-dominated universe with immense radiation pressure. Only extremely rare, large-amplitude perturbations can overcome this pressure. When such a perturbation re-enters the Hubble horizon, its collapse is relativistic and occurs on a timescale comparable to the horizon time. There is no multistreaming, no virialization, and no chance for the perturbation to disperse—it either collapses directly into a black hole or is erased by pressure.

    The collapse threshold is fully nonlinear and follows the Gaussian peaks model \cite{Polnarev2007}. Relativistic simulations show that PBHs form only if the density contrast exceeds $\delta_{\rm PBH} \sim 0.3$--$0.7$ (depending on the fluctuation profile) \cite{Carr1975,Polnarev2007,Escriva2020}. This is in stark contrast to the linear-theory threshold $\delta_c \approx 1.686$ used for halo collapse. Although the abundance expressions for PBHs may structurally resemble those of Press–Schechter, the underlying physical mechanisms differ fundamentally: PBH formation requires general relativity, a relativistic equation of state, horizon-scale dynamics, and strong nonlinearity. Halo formation is Newtonian, slow, and halted by violent relaxation.
    
    \item \textit{Relic black holes from a cosmological bounce:} In the BHU scenario, structure growth during a collapsing phase can produce overdensities or compact objects that lead to black holes after the bounce. This process, which we term \textit{Bounce Dark Matter} (BDM), operates via two channels:
    
\medskip
\noindent\textbf{Horizon-reentry channel.}
During the pre-bounce collapse, massive dark matter halos can form and virialize (with overdensities $\delta \approx 200$; see \S\ref{sec:virialization}). If such a region becomes larger than the particle horizon, it effectively “freezes out” (becoming causally disconnected) and survives as a self-gravitating cloud through the bounce. Upon re-expansion, this overdense region re-enters the horizon with $\delta \sim 200$, far above the threshold for relativistic collapse: $\delta_{\rm PBH} \sim 0.3$--$0.7$. It therefore collapses nearly instantaneously into a black hole upon horizon reentry. This mechanism closely parallels PBH formation at horizon entry, but in this case the initial overdensity results from gravitational instability rather than primordial quantum fluctuations.

\medskip
\noindent\textbf{Horizon-shielded channel.}
During collapse, a fraction of matter may undergo stellar evolution and form compact objects (e.g., black holes or neutron stars) before the bounce. These remnants behave as collisionless particles. If sufficiently larger than the particle horizon at bounce time, they pass through the hot bounce phase largely intact. Being encapsulated by their own event horizons (or simply nonrelativistic and non-interacting), they survive the bounce and reappear as \textit{relic} black holes once the horizon grows again in the expanding phase.

\end{enumerate}

Both BDM channels produce a relic black hole population spanning a broad mass range—from sub-solar to potentially supermassive. Whether these relics constitute a significant fraction of dark matter depends on their abundance, mass distribution, clustering, and survival across the bounce.

The BHU scenario even suggests a speculative yet intriguing possibility: our observable Universe may itself be a relic bounce structure formed during a previous cosmic cycle. In a larger parent cosmos lacking a cosmological constant, structure formation could proceed indefinitely. In such a scenario, extremely massive halos might form and collapse into regions that exit the Hubble horizon and undergo a bounce. These regions could become BHUs---self-contained universes within black hole event horizons---each following its own cosmological evolution. Many such relic BHUs (and smaller relic black holes) could arise in a single cosmic cycle. However, only the largest BHUs would evolve long enough to form galaxies, stars, and observers. Less massive BHUs would rapidly enter a de Sitter phase and “freeze out” before complexity can develop. In this sense, our Universe might be the interior of a rare, high-mass BHU relic born from a prior cosmic collapse.

A promising observational avenue to test the BDM idea lies in its potential imprint on the CMB. Both primordial and bounce-generated black holes would form on small scales (well below the galactic scale). Perturbations on these tiny scales re-enter the horizon well before recombination. Thus, their impact on the CMB would not appear in primary anisotropies (the acoustic peaks at $\ell \lesssim 3000$), which are insensitive to sub-Mpc structure due to Silk damping.

However, such early-forming black holes may induce subtle \textit{secondary} anisotropies on arcminute scales. For instance, compact dark matter objects could gravitationally lens CMB photons or produce a kinetic Sunyaev–Zel'dovich (kSZ) effect as they move through the photon-baryon fluid post-recombination. These effects could generate excess small-scale power beyond standard expectations. Upcoming high-resolution CMB experiments (e.g., ACT, Simons Observatory, CMB-S4) will probe multipoles up to $\ell \sim 3000$ and beyond. A detected excess of small-scale power (or anomalous lensing or kSZ signatures) could provide evidence of early compact objects, offering a new window into BHU physics and the Universe’s small-scale structure.

\section{Discussion and Conclusions}
We have proposed a new dark matter mechanism, \textit{Bounce Dark Matter} (BDM), in which relic black holes originate from a pre-Big-Bounce collapse phase. In summary, there are two BDM formation channels: (i) \textit{horizon-reentry BDM}, where massive dark matter halos from the collapse phase re-enter the horizon after the bounce and promptly collapse into black holes; and (ii) \textit{horizon-shielded BDM}, where compact objects (black holes or neutron stars) formed before the bounce survive it and reappear in the expanding phase. Both channels naturally produce a spectrum of \textit{relic black holes} spanning a wide range of masses—from substellar to intermediate and supermassive scales. These relics could constitute some or all of the dark matter and offer a novel explanation for the presence of early supermassive black holes (SMBHs).

An appealing aspect of the BDM scenario is that it addresses two fundamental problems within a single framework. First, it provides a non-particle dark matter candidate: an abundant population of primordial-origin black holes spread across a range of mass scales. These objects would behave as cold dark matter, clustering gravitationally while avoiding conflicts with baryon-based constraints (as they are encapsulated within event horizons). Second, BDM naturally offers \textit{seeds} for cosmic structure. The most massive relic black holes could reach $10^5$--$10^8\,M_{\odot}$, matching the requirements for SMBH seeds at high redshift. Even lower-mass relics ($\sim 10$--$100\,M_{\odot}$) could play a significant role through mergers or via early Hawking radiation (in the case of very low-mass relics).

Our proposal leads to several observational predictions. BDM generically yields a broad black hole mass spectrum rather than a monochromatic one. If a significant fraction of dark matter resides in black holes of tens of solar masses, this could enhance the merger rate detected by gravitational-wave observatories. The LIGO/Virgo detections of binary black hole mergers may thus include contributions from a relic BDM population, whose spatial clustering or formation history differs from stellar-origin black holes. Microlensing surveys and dynamical constraints offer further avenues to probe compact dark matter: future observations of stellar microlensing, astrometric shifts, or CMB distortions could detect signatures consistent with BDM predictions. At the high-mass end, the existence of billion-solar-mass quasars at $z > 7$ could be explained if their progenitors were relic black holes of $M \sim 10^5$--$10^6\,M_{\odot}$ formed before the bounce, alleviating the need for unusually rapid post-Big-Bang SMBH growth.

Another intriguing implication is the potential to infer pre-bounce conditions from present-day observations. If a significant fraction of dark matter consists of relic black holes, one can infer the perturbation spectrum and collapse dynamics of the prior cosmic cycle. During collapse, linear density perturbations grow as $\delta \propto a^{-3/2}$ as $a \to 0$, much faster than in an expanding universe. This means that modest initial fluctuations can grow into highly overdense regions by the time of the bounce, naturally producing the conditions for BDM. Conversely, matching the current dark matter density imposes constraints on the amplitude of perturbations and the timescale of collapse before the bounce. In this way, relic black holes offer a unique observational window into the pre-Big-Bang universe.

Beyond dark matter, gravitational waves, and early black holes, the BHU scenario may yield additional testable consequences. A finite collapsing universe could carry small amounts of angular momentum or anisotropy, potentially leaving imprints as rotational B-modes in the cosmic microwave background or 
weak gravitational lensing, galaxy spin alignments \cite{Shamir2025}
or tension in the $H_0$ measurements
\cite{2025MNRAS.538.3038S}. While torsion-based modify gravity bounce models \cite{Poplawski2016} explicitly introduce rotation to trigger the bounce, in the BHU framework such rotation could arise naturally from initial conditions within classical General Relativity (see Appendix in \cite{Gaztanaga2022a}). Though these effects remain speculative, they highlight the rich phenomenology of bounce cosmologies. The BHU scenario offers a consistent mechanism for early- and late-time acceleration, addresses the dark matter problem, and provides a novel perspective on cosmic initial conditions—all within the standard framework of gravity.

\begin{acknowledgments}
E.G. acknowledges support from the Spanish Ministry of Science and Innovation project PGC2018-102021-B-100 and the Maria de Maeztu Unit of Excellence (Institute of Space Sciences, CEX2020-001058-M).
\end{acknowledgments}

\bibliographystyle{apsrev4-2}
\bibliography{BounceDM}

\appendix 

\section{The FLRW* cloud perturbation}
\label{sec:perturbation}

Here we consider relativistic spherical perturbations that are uniform (a FLRW cloud) within an approximately empty background. The background may be a lower-density FLRW or a compensated perturbation, as illustrated in Fig.~\ref{fig:collapse}. This setup can model a spherical collapse perturbation within our universe, or alternatively, model our entire universe as a spherical collapse within a larger background---the so-called BHU scenario.

In the Newtonian gauge, a FLRW perturbation takes the form:
\beq
ds^2 = -(1+2\Psi)\,dt^2 + \frac{dr^2}{1+2\Phi} + r^2 d^2\Omega,
\label{eq:Newtoniagauge}
\eeq
where the areal radius $r$ can also be written in terms of the comoving radius $\chi$: $r= a(\tau) \chi$. Note how this metric has the same form as the most general spherically symmetric background. This shows that the same metric form used to model a global background can also model a local spherical perturbation.

In the \textit{comoving gauge} with spherical symmetry:
\beq
ds^2 = -d\tau^2 + \frac{a^2\,d\chi^2}{1-K(r)\chi^2} + a^2\chi^2 d^2\Omega,
\label{eq:Comoving-gauge}
\eeq
where $K(r)$ is the \textit{curvature profile} \cite{Polnarev2007}, and $\Phi=-\frac{1}{2} K(r) r^2$ is the metric potential. The time $\tau$ is set such that $g_{00}=1$ and $g_{0\mu}=0$; this defines a synchronous frame in which worldlines at fixed spatial coordinates follow geodesics.

For a uniform overdensity, the curvature is constant $K=+1/\chi_k^2$, reproducing the closed-FLRW geometry used in the \textit{BHU model} \cite{Gaztanaga2025}. This corresponds to a local (finite) FLRW metric, which can be interpreted as a global solution to Einstein's field equations with appropriate junctions \cite{BH_interior_Stuckey,Geller2018,gaztanaga:bhu1}. Specifically, a closed FLRW interior is matched to a FLRW exterior, a Schwarzschild exterior, or a compensated perturbation that includes both (see Fig.~\ref{fig:collapse}). The finite perturbation has a finite mass $M$:
\beq
r_s = 2GM = \frac{8\pi G }{3} \rho_0 \, \chi_*^3,
\eeq
where $\rho_0$ is the density at $a=1$ (when the perturbation forms), and $\chi_*$ is the constant comoving radius of the junction in a matter-dominated case. 

If the perturbation follows 
the FLRW solution with a general equation of state $P= P(\rho)$, we have:
\bea
H^2 =  \left(\frac{\dot{a}}{a} \right)^2 &=& \frac{8\pi G}{3} \rho -\frac{k}{a^2}\,+\frac{\Lambda}{3},
\label{eq:Hubble-LTB3} \\
\frac{\ddot{a}}{a} = \frac{\Lambda}{3} - \frac{4\pi G}{3} (\rho + 3P)  &;& 
 \dot\rho =- 3~ H~(\rho+P).
\label{eq:ddota}
\eea
Even if the background to our perturbation contains a $\Lambda$ term, it's not obvious that the perturbation should include one initially. If $\Lambda$ is a fundamental constant, it might appears in both background and perturbation equations. If it's an effective dark energy fluid, its contribution to $H^2$ depends on the time of perturbation formation.

To simplify, we assume a single fluid with $P=\omega \rho$. At $a=1$ (formation), we take $\Lambda$ to be negligible. For a matter-dominated cloud ($\omega \simeq 0$), the closed-FLRW solution has a turnaround point $a=a_{\rm max}$. The turnaround radius exceeds the Schwarzschild radius: $a_{\rm max} \chi_*>r_s$. As collapse proceeds ($a \to 0$), the perturbation enters its event horizon: $a\chi_*<r_s$, unless pressure intervenes earlier. For a very large cloud ($M \simeq 5\times 10^{22}\, M_\odot$ \cite{gaztanaga_mou}), the density at horizon crossing is low, and pressure (including dark matter virialization) can be neglected. For smaller masses, baryonic or dark matter pressure delays BH formation (see \S\ref{sec:virialization}).

\section{The Black Hole Universe (BHU)}
\label{sec:BHU}

Early precursors of the BHU concept include spherical models with a finite, homogeneous FLRW* patch embedded in empty space (for a review, see \cite{Faraoni2015}). These mirror the Newtonian spherical collapse model \cite{Faraoni2020}. Examples include Lema\^itre (1933) \cite{Lemaitre:1933gd}, Tolman (1934) \cite{Tolman1934}, and Oppenheimer \& Snyder (1939) \cite{Oppenheimer1939}, which describe collapse using a pressureless FLRW interior matched to a Schwarzschild exterior. Further developments by Bondi \cite{Bondi1947}, Misner \& Sharp \cite{Misner1964}, Vaidya \cite{Vaidya1968}, and Hawking \cite{Hawking1968} maintained this framework. These models admit expansion and collapse but result in singularities \cite{Hawking1970,HawkingEllis1968}, and were designed for stellar collapse, not cosmology.

In the 1970s, Pathria \cite{Pathria1972} and Good \cite{Good1972} suggested our Universe could reside inside a black hole. Zhang (2018) \cite{Zhang2018} and others explored related ideas, but lacked exact GR solutions. More recent developments include nonsingular bounces in modified gravity \cite{Poplawski2016,Brandenberger2017}, though these are not strictly BHU.

In 1994, Stuckey \cite{BH_interior_Stuckey} introduced an exact GR solution for expansion inside a BH, but it still had singularities. A full BHU model including $p \neq 0$ was developed in \cite{Gaztanaga2021,Gaztanaga2022a,gaztanaga:bhu1,Gaztanaga2022c,Gaztanaga2025}. Earlier versions ($k \simeq 0$) focused on late-time acceleration; for bounce cosmology, curvature ($k=+1$) is essential as it describes a closed, finite cloud.

As discussed in \cite{Gaztanaga2025}, high-density collapse yields a stiff equation of state. The energy density reaches a constant ground value $\rho \to \rho_G$ as $a \to a_B \ll 1$, providing quantum \textit{degeneracy pressure}. At this point, $\dot\rho = 0$ and $\omega = -1$ (from Eq.~\ref{eq:ddota}), saturating the null energy condition (NEC): $T_{\mu\nu}k^\mu k^\nu = 0$.

Penrose’s singularity theorem requires a non-compact Cauchy surface, which closed FLRW models lack. With NEC saturation, a gravitational bounce becomes possible.
Using the FLRW equations (Eqs.~\ref{eq:Hubble-LTB3}–\ref{eq:ddota}) with $\rho = \rho_G$ and $P = -\rho_G$:
\bea
\frac{\ddot{a}}{a} &=& +\frac{8\pi G}{3}\,\rho_G
          \equiv \frac{r_S}{R_G^3},
\qquad \dot\rho_G = 0,\\[3pt]
H^2 &=& \left(\frac{\dot a}{a}\right)^2
       = \frac{r_S}{R_G^3} - \frac{k}{a^2},
\label{eq:Hubble-LTB4}
\eea
where $R_G$ is the ground-state radius and $r_S = 2GM$ is the Schwarzschild radius.

The bounce occurs when $\dot a = 0$, $\ddot a > 0$, yielding:
\beq
a_B = \sqrt{\frac{R_G^3}{\chi_k^2\,r_S}},
\qquad
R_B^2 = \frac{R_G^3}{r_S},
\label{eq:a_B}
\eeq
valid only for $k = +1$. The exact solution is:
\beq
\frac{a}{a_B} = \cosh\!\left(\frac{\tau}{R_B}\right)
             = \tfrac{1}{2}\!\left[e^{+\tau/R_B} + e^{-\tau/R_B}\right],
\label{eq:bounce}
\eeq
with $R_B = a_B\chi_k = \sqrt{3/(8\pi G\rho_G)}$. This describes pre-bounce contraction ($\tau < 0$) and post-bounce inflationary expansion ($\tau > 0$).
This inflationary phase solves the horizon and flatness problems \cite{Starobinskii1979,Guth1981,Albrecht1982,Linde1982}. 

Cosmic acceleration today ($\ddot{a} > 0$) is attributed to $\Lambda$, interpreted either as a fundamental constant $\Lambda_\text{F}$ or effective fluid $\Lambda_\text{DE}$, with $\rho_G \gg \rho_\text{DE}$. Its length scale $R_\Lambda = \sqrt{3/\Lambda}$ far exceeds $R_G$, so $\Lambda$ can be neglected during the bounce.

However, after the bounce, the BHU model interprets $\Lambda$ as a boundary effect from the collapsing cloud’s gravitational radius. Since $R_\Lambda = r_S = 2GM$, this gives:
\beq
\Lambda = \frac{3}{r_S^2},
\eeq
so $\Lambda$ reflects the total mass $M$ of the Universe---a natural, intuitive origin. More detailed derivations are given in \cite{gaztanaga:bhu1} using Israel matching and extrinsic curvature.

In the BHU framework, the bounce can be described within classical General Relativity as a closed FLRW overdensity whose effective equation of state evolves from dust-like matter to a quantum-supported ground state of nearly constant density, producing a quasi--de~Sitter phase. The underlying justification is quantum: during collapse, the Pauli exclusion principle drives matter into a highly degenerate state with effective negative pressure, halting contraction and triggering re-expansion. A complete picture therefore involves two successive transitions—classical$\rightarrow$quantum before the bounce and quantum$\rightarrow$classical after it—both occurring near a quasi--de~Sitter geometry. The macroscopic arrow of time does not reverse, since it is defined locally by entropy flow rather than by global spacetime evolution. Near the bounce, the particle horizon shrinks to microscopic scales, and the quantum state naturally contains both contracting and expanding PT-conjugate components. The decoherence-driven emergence of classicality during expansion underlie the observed parity (PT) asymmetry in primordial fluctuations: CMB modes correspond to perturbations that became classical after the bounce, where expansion effectively breaks time-reversal symmetry and can imprint an excess of odd-parity power \cite{Gaztañaga_2026}. In this scenario, relic perturbations add to the standard scale-invariant spectrum of inflated quantum fluctuations; they are subdominant on large scales (since large masses are exponentially suppressed) but could dominate on smaller scales. Consequently, PBH constraints from CMB $\mu$-distortions, which assume Gaussian scale-invariant primordial fluctuations, do not necessarily apply to these non-Gaussian relic contributions.

\subsection{Junction for $P \ne 0$ and $\Lambda \ne 0$}

In the matter-dominated case ($P = 0$), the junction between a FLRW cloud and a Schwarzschild background occurs at constant $\chi = \chi_*$ \cite{BH_interior_Stuckey,gaztanaga:bhu1}. With pressure, \cite{Lemaitre:1933gd,Faraoni2015} give:
\begin{equation}
\dot M = -4\pi G P R^2 \dot R.
\end{equation}
To maintain constant $M$ (from the external Schwarzschild view), we need $\dot R \to 0$, implying:
\begin{equation}
\chi_* \to \chi_0 e^{- \int H d\tau} = \chi_0 a^{-1}.
\label{eq:rsmatch}
\end{equation}

In the expanding case with $\Lambda$, this occurs if the junction $\chi_*$ traces a future null surface of the FLRW cloud:
\begin{equation}
\chi_* = \int \frac{d\tau}{a} = \int_a^\infty \frac{da}{H a^2} \to R_\Lambda a^{-1},
\end{equation}
as $a \to \infty$, $H \to R_\Lambda^{-1}$, so $\chi_* \to r_\Lambda a^{-1}$ and $R \to r_\Lambda = r_s$. This completes the BHU picture: a FLRW interior inside a Schwarzschild black hole.

Near the bounce ($\dot \rho \to 0$, $\omega \to -1$), we again find $\chi_* \to R_\Lambda a^{-1}$, $\dot R \to 0$, and constant $r_S$, satisfying the null junction condition \cite{gaztanaga:bhu1}.

\end{document}